\documentclass{article}
\usepackage{mlspconf,amsmath,graphicx}

\usepackage{amssymb, latexsym}
\usepackage{amsmath} 
\usepackage{amsthm}
\usepackage{bm}
\usepackage{graphicx}
\usepackage{epstopdf}  
\usepackage[mathscr]{eucal}
\usepackage{enumerate}
\usepackage{cite}
\usepackage[font=small,skip=1pt]{caption}
\usepackage{subfig}
\usepackage{float}
\usepackage{multirow}
\usepackage{array}
\usepackage[dvipsnames]{xcolor}

\usepackage{hyperref}

\graphicspath{{./figures/}{./}}

\title{Semi-supervised source localization with deep generative modeling}
\name{Michael J. Bianco$^1$, Sharon Gannot$^2$, and Peter Gerstoft$^1$
\thanks{This work is supported by the Office of Naval Research, Grant No. N00014-11-1-0439 and the European Union’s Horizon 2020 Research and Innovation Programme under Grant Agreement No. 871245.}}
\address{${}^1$ Marine Physical Laboratory, UCSD, CA, USA, mbianco@ucsd.edu \\
${}^2$ Faculty of Engineering, Bar-Ilan University, Israel}

\begin{document}
%\ninept
%
\maketitle
\begin{abstract}
We propose a semi-supervised localization approach based on deep generative modeling with variational autoencoders (VAEs). Localization in reverberant environments remains a challenge, which machine learning (ML) has shown promise in addressing. Even with large data volumes, the number of labels available for supervised learning in reverberant environments is usually small. We address this issue by performing semi-supervised learning (SSL) with convolutional VAEs. The VAE is trained to generate the phase of relative transfer functions (RTFs), in parallel with a DOA classifier, on both labeled and unlabeled RTF samples. The VAE-SSL approach is compared with SRP-PHAT and fully-supervised CNNs. We find that VAE-SSL can outperform both SRP-PHAT and CNN in label-limited scenarios.
%Further, the VAE-SSL generated RTF phase patterns are assessed.
\end{abstract}
\begin{keywords}
Source localization, semi-supervised learning, generative modeling, deep learning
\end{keywords}
%
%\today
\section{Introduction}
\label{sec:intro}

Source localization is an important problem in acoustics and many related fields. The performance of localization algorithms is degraded by reverberation, which induces complex temporal arrival structure across sensor arrays. Despite recent advances, e.g. \cite{purwins_deep_2019,bianco2019machine,gannot2019introduction}, acoustic localization in reverberant environments remains a major challenge. 
Recently, there has been great interest in machine learning (ML)-based techniques  in acoustics, including source localization and event detection \cite{vincent2018audio,mesaros2017dcase,chakrabarty2019multi,adavanne2019sound,ping2020three,ozanich2020feedforward}. A challenge for ML-based methods in acoustics is the limited amount of labeled data and the complex acoustic propagation in natural environments, despite  large volumes of recordings\cite{purwins_deep_2019,bianco2019machine}. This limitation has motivated recent approaches for localization based on semi-supervised learning (SSL)\cite{laufer2016semi,opochinsky2019deep}. In SSL, ML models are trained using many examples with only few labels, with goal of exploiting the natural structure of the data.

We propose an SSL localization approach based on deep generative modeling with variational autoencoders (VAEs)\cite{kingma2014auto}. Deep generative models \cite{goodfellow2016deep}, e.g. generative adversarial networks (GANs)\cite{goodfellow2014generative}, have received much attention for their ability to learn high-dimensional sample distributions, including those of natural images \cite{karras2019analyzing}. GANs in acoustics have had success in generating raw audio \cite{oord2016wavenet} and speech enhancement \cite{donahue2018exploring}. VAEs learn from unlabeled data explicit latent codes for generating samples, and are inspiring examples of representation learning\cite{kingma2014semi,bengio2013representation}.

We perform SSL based on VAEs to encode the phase of the relative transfer function (RTF) \cite{gannot2001signal} between two microphones to a latent parametric distribution. The resulting model estimates DOA and generates RTF phase. By learning to generate RTF phase, the VAE-SSL system learns implicit physics from unlabeled samples. The VAE-SSL method is implemented using convolutional neural networks (CNNs). The performance of the convolutional VAE-SSL in reverberant environments is assessed against the steered response power with phase transform (SRP-PHAT)\cite{brandstein1997robust}, and supervised convolutional neural network (CNN) approaches.

\section{Theory}
We use RTFs\cite{gannot2001signal}, specifically the RTF phase, as the acoustic feature for our VAE-SSL approach. The RTF is independent of the source and well-represents the physics of the acoustic system. This helps to focus ML on physically relevant features. We encode the RTF phase as a function of source azimuth (direction of arrival, DOA).

\subsection{Relative transfer function (RTF)}
We consider time domain acoustic recordings of the form
\begin{align}\label{eq:mics}
d_i&=a_i*s+u_i,
\end{align}
with $s$  the source signal $a_i$ ($i=\{1,2\}$ the microphone index) the impulse responses (IRs) relating the source and each of the microphones, $u_i$ noise signals which are independent of the source.
Define the acoustic transfer functions $A_i(k)$ as the Fourier transform of the IRs $a_i$. Then, the relative transfer function (RTF) is defined as\cite{gannot2001signal}
\begin{equation}
    H(k) = \frac{A_2(k)}{A_1(k)},
\end{equation}
with $k$ the frequency index. With $d_1$ as reference, $H(k)$ is
% %
% \begin{align}\label{eq:mics_rtf_fft2}
% \widehat{H}(k)=\frac{A_2(k)}{A_1(k)}=\frac{S_{d_2d_1}}{S_{d_1d_1}}, 
% \end{align}
% %
%
\begin{align}\label{eq:mics_rtf_fft2}
\widehat{H}(k)=\frac{S_{d_2d_1}}{S_{d_1d_1}}, 
\end{align}
with $S_{d_1d_1}=D_1(k)^*D_1(k)$ the power spectral density (PSD), $S_{d_2d_1}=D_2(k)^*D_1(k)$ the cross-PSD for a single frame and $\cdot^{*}$  the complex conjugate. This estimator is biased since we neglect the PSD of the noise. An unbiased estimator can be obtained, but we observe here that the biased estimate $\widehat{H}(k)$ works well, see also \cite{markovich2018performance,koldovsky2015spatial}. For each FFT frame, a vector of RTFs is obtained $\widehat{\mathbf{h}}=[\widehat{H}(1)\dots \widehat{H}(K)]^\mathrm{T}\in\mathbb{C}^K$ for $K$ frequency bins. We  use RTFs estimated using a single frame as input features to the VAE-SSL.

The $n\text{th}$ input sample to VAE-SSL and the supervised CNN is a sequence of RTF frames $\mathbf{x}_n=\text{vec}(\text{phase}(\widehat{\mathbf{H}}_n))\in\mathbb{R}^{KP}$, with $\widehat{\mathbf{H}}_n=[\widehat{\mathbf{h}}_n\dots \widehat{\mathbf{h}}_{n+P-1}]\in\mathbb{C}^{K\times P}$, $K=N_\mathrm{FFT}/2$, and $P$ the number of RTF frames in the sequence.

\subsection{Semi-supervised learning with VAEs}
We assume the wrapped RTF phase $\mathbf{x}$ is generated (disregarding subscript $n$ to simplify notation) by a random process involving the latent random variable $\mathbf{z}\in\mathbb{R}^M$ and source location label $\mathbf{y}\in\mathbb{R}^T$ with $T$ the number of candidate DOAs. We formulate a principled semi-supervised learning framework based on VAEs, which treats the label $\mathbf{y}$ as both latent and observed, and trains a classifier using both labeled and unlabeled data. This corresponds to the `M2' model in \cite{kingma2014semi}. Starting with Bayes' rule we have for labeled data

\begin{align}\label{eq:bayes2}
p_\theta(\mathbf{z}|\mathbf{x},\mathbf{y})=\frac{p_\theta(\mathbf{x},\mathbf{y}|\mathbf{z})p_\theta(\mathbf{z})}{p_\theta(\bf{x,y})}
\end{align}
and for unlabeled data
\begin{align}\label{eq:bayes3}
p_\theta(\mathbf{y},\mathbf{z}|\mathbf{x})=\frac{p_\theta(\mathbf{x}|\mathbf{y},\mathbf{z})p_\theta(\mathbf{y},\mathbf{z})}{p_\theta(\bf{x})}
\end{align}
Using \eqref{eq:bayes2} as an example, direct estimation of the posterior $p_\theta(\mathbf{z}|\mathbf{x},\mathbf{y})$ is nearly always intractable due to $p(\mathbf{x},\mathbf{y})=\int p(\mathbf{x},\mathbf{y},\mathbf{z})d\bf{z}$. As will later be shown, the subscript $\theta$ indicates distributions and functions defined using the decoder network, i.e. the generative model in the VAE.

\begin{figure}[t]
\centering
\includegraphics[width = 3in]{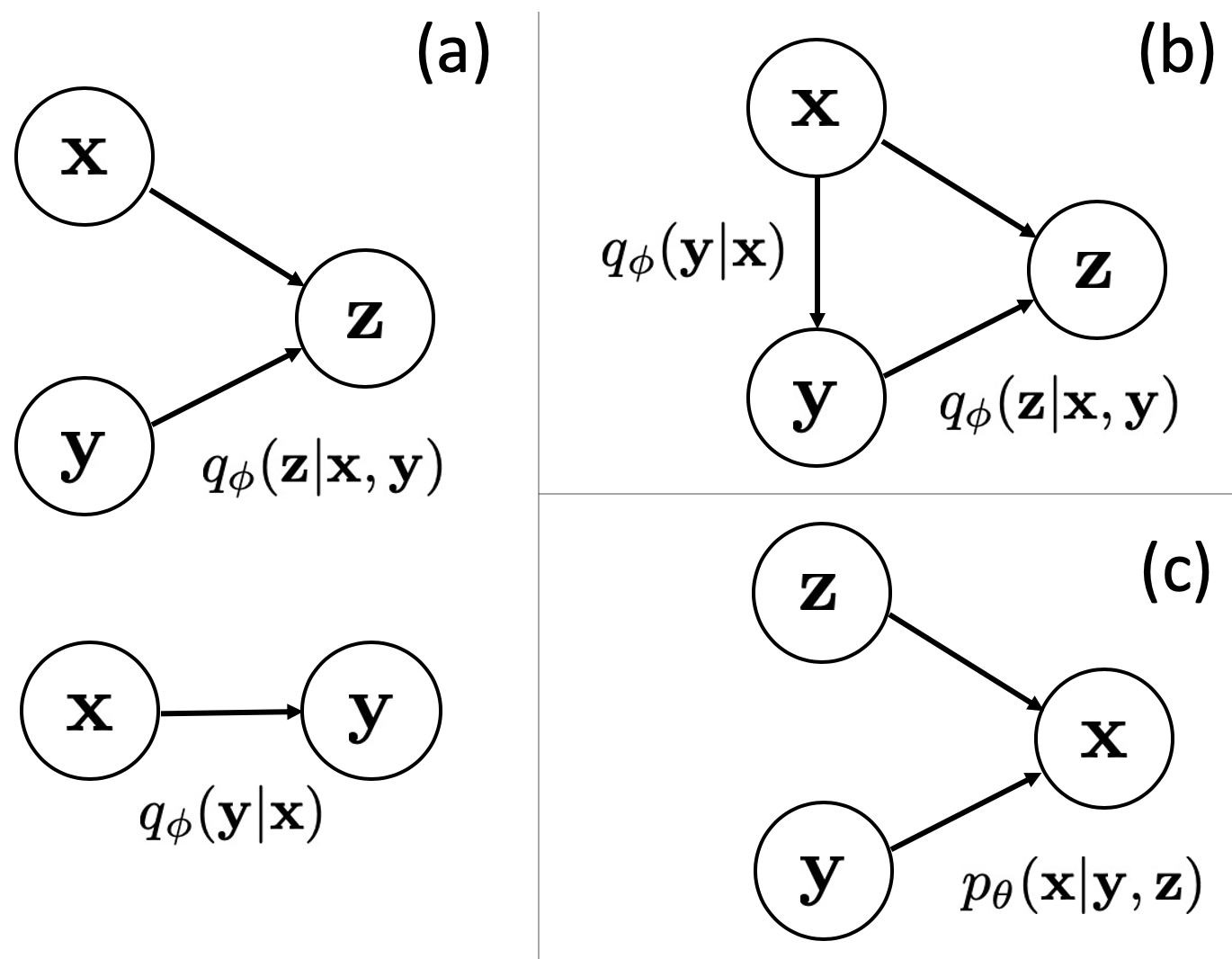}
\caption{Directed graphical model for semi-supervised learning with VAE. (a) Inference models (encoders), supervised learning (labeled) per \eqref{eq:label_k2},\eqref{eq:label_k6}. (b) Inference models, unsupervised (unlabeled) per \eqref{eq:label_k4}. (c) Generative model (decoder).}
\label{fig:ssvvae_dgm}
\end{figure}

VAEs\cite{kingma2014semi,kingma2019introduction} approximate posterior distributions using variational inference (VI)\cite{blei2017variational}, a family of methods for approximating conditional densities which relies on optimization instead of (MCMC) sampling. In VAEs the conditional densities are modeled with NNs. A variational approximation to the intractable posterior $p_\theta(\mathbf{z}|\mathbf{x},\mathbf{y})$ is defined by the encoder network as $q_\phi(\mathbf{z}|\mathbf{x},\mathbf{y})\approx p_\theta(\mathbf{z}|\mathbf{x},\mathbf{y})$, with the subscript $\phi$ corresponding to distributions and functions defined using the encoder network. The networks constituting the VAE-SSL model are shown in Fig.~\ref{fig:ssvvae_dgm}.

Starting with the posterior for the labeled data (see \eqref{eq:bayes2}), per VI we seek $q_\phi(\mathbf{z|x,y})$ which minimizes the KL-divergence
\begin{align}\label{eq:kl2}
\{\phi,\theta\}=\underset{\phi,\theta}{\arg\min}~\mathrm{KL}(q_\phi(\mathbf{z|x,y})||p_\theta(\mathbf{z|x,y})).
\end{align}
The intractable posterior is approximated by $q_\phi(\mathbf{z}|\mathbf{x},\mathbf{y})\approx p_\theta(\mathbf{z}|\mathbf{x},\mathbf{y})$. Assessing the $\mathrm{KL}$-divergence, we obtain
\begin{align}\label{eq:label_kl}
\mathrm{KL}&(q_\phi(\mathbf{z}|\mathbf{x},\mathbf{y})||p_\theta(\mathbf{z}|\mathbf{x},\mathbf{y})) \nonumber \\
&=\mathbb{E}\big[\log q_\phi(\mathbf{z}|\mathbf{x},\mathbf{y})\big]-\mathbb{E}\big[\log p_\theta(\mathbf{x},\mathbf{y}|\mathbf{z})p_\theta(\mathbf{z})\big] \\ \nonumber
&\hspace{5ex} + \log p_\theta(\mathbf{x,y}) \\ \nonumber
&=-\mathrm{ELBO}+ \log p_\theta(\mathbf{x,y}),
\end{align}
with the expectation $\mathbb{E}$ relative to $q_\phi(\mathbf{z}|\mathbf{x},\mathbf{y})$. This reveals the dependence of the $\rm KL$ divergence on evidence $p_\theta(\mathbf{x,y})$, which is intractable. The other two terms in \eqref{eq:label_kl} form the evidence lower bound (ELBO). Since the KL divergence is non-negative, the ELBO `lower bounds' the evidence: $\text{ELBO}\le\log p_\theta(\mathbf{x,y})$. Maximizing the ELBO is equivalent to minimizing the $\mathrm{KL}$ \eqref{eq:kl2}. We thus minimize $-\text{ELBO}$. 

Considering the ELBO terms from \eqref{eq:label_kl}, we formulate the objective for labeled data, assuming $\mathbf{y}$ and $\mathbf{z}$ independent in the generative model,
\begin{gather}
\begin{aligned}
-C(\theta,\phi;\mathbf{x,y})&=\mathbb{E}\big[\log p_\theta(\mathbf{x},\mathbf{y}|\mathbf{z})p_\theta(\mathbf{z})-\log q_\phi(\mathbf{z}|\mathbf{x},\mathbf{y})\big] \\
&=\mathbb{E}\big[\log p_\theta(\mathbf{x}|\mathbf{y},\mathbf{z}) + \log p_\theta(\mathbf{y})+ \log p_\theta(\mathbf{z}) \\
& \hspace{5ex} -\log q_\phi(\mathbf{z}|\mathbf{x},\mathbf{y})\big].
\end{aligned} \raisetag{0.8\baselineskip}
\label{eq:label_k2}
\end{gather}

Next, an objective for unlabeled data is derived. The intractable posterior $q_\phi(\mathbf{y},\mathbf{z}|\mathbf{x})\approx p_\theta(\mathbf{y},\mathbf{z}|\mathbf{x})$ from \eqref{eq:bayes3}. From the $\mathrm{KL}$, we find the objective (negative ELBO) as
\begin{gather}
\begin{aligned}
-D(\theta,\phi;\mathbf{x})&=\mathbb{E}\big[\log p_\theta(\mathbf{x}|\mathbf{y},\mathbf{z})p_\theta(\mathbf{y},\mathbf{z})-\log q_\phi(\mathbf{z},\mathbf{y}|\mathbf{x})\big] \\
&=\mathbb{E}\big[\log p_\theta(\mathbf{x}|\mathbf{y},\mathbf{z})+\log p_\theta(\mathbf{y})+\log p_\theta(\mathbf{z})\\
& \hspace{5ex} -\log q_\phi(\mathbf{y}|\mathbf{x})-\log q_\phi(\mathbf{z}|\mathbf{x})\big]
\end{aligned} \raisetag{0.8\baselineskip}
\label{eq:label_k3}
\end{gather}
with the expectation relative to $q_\phi(\mathbf{z},\mathbf{y}|\mathbf{x})$. Further expanding \eqref{eq:label_k3} we obtain
\begin{gather}
\begin{aligned}
-D(\theta,&\phi;\mathbf{x})=\mathbb{E}_{q_\phi(\mathbf{y}|\mathbf{x})}\big[ \mathbb{E}_{q_\phi(\mathbf{z}|\mathbf{x})} \big[ \log p_\theta(\mathbf{x}|\mathbf{y},\mathbf{z})+\log p_\theta(\mathbf{y})\\
& \hspace{5ex} +\log p_\theta(\mathbf{z})-\log q_\phi(\mathbf{y}|\mathbf{x})-\log q_\phi(\mathbf{z}|\mathbf{x})\big]\big] \\
&=\sum_\mathbf{y}q_\phi(\mathbf{y}|\mathbf{x})\big[-C(\theta,\phi;\mathbf{x,y})-\log q_\phi(\mathbf{y}|\mathbf{x})\big]
\end{aligned} \raisetag{1.0\baselineskip}
\label{eq:label_k4}
\end{gather}

An overall objective for training using labelled and unlabeled data is derived by combining \eqref{eq:label_k2},\eqref{eq:label_k4} as
\begin{align}\label{eq:label_k5}
\mathcal{J} = \sum_{\{\mathbf{x,y}\}} C(\theta,\phi;\mathbf{x,y}) + \sum_{\{\mathbf{x}\}} D(\theta,\phi;\mathbf{x}).
\end{align}
%

%%%%%%%%%%%%%%%%%%%%%%%%%%%%%%%
%% TABLE 1 %%%%%%%%%%%%%%%%%%%
%%%%%%%%%%%%%%%%%%%%%%%%%%%%%%%
\begin{table}[t]
\caption{Design case: MAE ($^\circ$) and Accuracy (\%) of VAE-SSL and alternative approaches. SRP-PHAT results in bottom row.}\label{table:exp0}
  \centering
  \begin{tabular}{c | l | l | l | l }
    \cline{2-5}
     \multicolumn{1}{c|}{} & \multicolumn{2}{c|}{VAE-SSL} & \multicolumn{2}{c}{CNN}\\ \hline
     $J$ (\# labels) & MAE & Acc. & MAE & Acc.\\ \hline\hline
    37 & 27.2 & 9.59 & 39.5 & 13.4 \\
    74 & 10.7 & 36.0 & 33.3 & 21.3 \\
     148 & 13.7 & 39.5 & 27.3 & 24.7 \\
    481 & 4.37  & 69.8 & 8.32 & 59.4 \\
    999 & 2.24  & 83.7 & 0.34 & 96.4 \\
     \hline\hline
     \multicolumn{1}{c|}{SRP-PHAT} & \multicolumn{1}{l|}{18.0} & \multicolumn{1}{l|}{11.6} & \multicolumn{2}{c}{}\\ 
     \cline{1-3}
  \end{tabular}
\end{table} 
%%%%%%%%%%%%%%%%%%%%%%%%%%%%%%%
%% TABLE 2 %%%%%%%%%%%%%%%%%%%
%%%%%%%%%%%%%%%%%%%%%%%%%%%%%%%
\begin{table}[t]
\caption{Validation case: Notation as Table~\ref{table:exp0}.
%MAE (deg.) and Accuracy (\%) of VAE-SSL and alternative approaches. SRP-PHAT results in bottom row.
}\label{table:exp1}
  \centering
  \begin{tabular}{c | l | l | l | l }
    \cline{2-5}
     \multicolumn{1}{c|}{} & \multicolumn{2}{c|}{VAE-SSL} & \multicolumn{2}{c}{CNN}\\ \hline
     $J$ (\# labels) & MAE & Acc. & MAE & Acc.\\ \hline\hline
    37  & 29.2  & 9.03  & 49.6  & 10.2 \\ 
    74	& 14.5	& 31.5	& 43.8	& 17.0 \\
    148	& 17.2	& 36.2	& 34.6	& 21.4\\
    481	& 7.00	& 65.9	& 10.2	& 56.9\\
    999	& 5.48  & 74.0	& 1.32	& 90.5\\
     \hline\hline
     \multicolumn{1}{c|}{SRP-PHAT} & \multicolumn{1}{l|}{21.7} & \multicolumn{1}{l|}{12.3} & \multicolumn{2}{c}{}\\ 
     \cline{1-3}
  \end{tabular}
\end{table} 

Assessing the terms in \eqref{eq:label_k2}, the supervised learning objective $C$ does not condition the label $\mathbf{y}$ on the sample $\mathbf{x}$. The term $\log q_\phi(\mathbf{y}|\mathbf{x})$ is only present in the unsupervised learning objective \eqref{eq:label_k5}. This issue is remedied by an additional term
\begin{align}\label{eq:label_k6}
\mathcal{J}^\alpha = \mathcal{J} + \alpha \sum_{\{\mathbf{x,y}\}} -\log q_\phi(\mathbf{y}|\mathbf{x}).
\end{align}

It is assumed that the data is explained by the generative process (see \eqref{eq:label_k4} for terms): $p_\theta({\mathbf{y}})=\mathrm{Cat}(\mathbf{y}|\boldsymbol{\pi})$, with $\mathrm{Cat}(\cdot|\cdot)$ the categorical (multinomial) distribution and $\boldsymbol{\pi}\in\mathbb{R}^T$ the probabilities of the classes (which are assumed equal, and normalized such that $\sum_T\pi_t=1$); $p_\theta(\mathbf{z})=\mathcal{N}(\mathbf{0},\mathbf{I})$ (as before); and $p_\theta(\mathbf{x}|\mathbf{y},\mathbf{z})=\mathcal{N}(\mathbf{x}|\boldsymbol{\mu}_\theta(\mathbf{y,z}),\mathbf{I})$, with $\boldsymbol{\mu}_\theta(\mathbf{y,z})$ the decoder. The densities of the inference model are: $q_\phi(\mathbf{z}|\mathbf{y,x})=\mathcal{N}(\mathbf{z}|\boldsymbol{\mu}_\phi(\mathbf{x,y}),\mathrm{diag}(\boldsymbol{\sigma}_\phi^2(\mathbf{x,y})))$, with $\boldsymbol{\mu}_\phi(\mathbf{x,y})$ and $\boldsymbol{\sigma}_\phi^2(\mathbf{x,y})$ the outputs of the z-encoder; and $q_\phi(\mathbf{y|x})=\mathrm{Cat}(\mathbf{y}|\boldsymbol{\pi}_\phi(\mathbf{x}))$, with $\boldsymbol{\pi}_\phi(\mathbf{x})$ the classifier network.

%%%%%%%%%%%%%%%%%%%%%%%%%%%%%%%
%% TABLE 3 %%%%%%%%%%%%%%%%%%%
%%%%%%%%%%%%%%%%%%%%%%%%%%%%%%%
\begin{table}[t]
\caption{Test case I:  Notation as Table~\ref{table:exp0}.
%MAE ($^\circ$) and Accuracy (\%) of VAE-SSL and alternative approaches. SRP-PHAT results in bottom row.
}\label{table:exp2}
  \centering
  \begin{tabular}{c | l | l | l | l }
    \cline{2-5}
     \multicolumn{1}{c|}{} & \multicolumn{2}{c|}{VAE-SSL} & \multicolumn{2}{c}{CNN}\\ \hline
     $J$ (\# labels) & MAE & Acc. & MAE & Acc.\\ \hline\hline
    37  & 27.3  & 10.3  & 43.9 & 10.6 \\
    74	& 11.0	& 33.2	& 37.2 & 17.6 \\
    148	& 13.7	& 37.1	& 29.2 & 22.4 \\
    481	& 5.78	& 66.0	& 10.5 & 54.1 \\
    999	& 2.97	& 76.8	& 0.57 & 93.2 \\
     \hline\hline
     \multicolumn{1}{c|}{SRP-PHAT} & \multicolumn{1}{l|}{20.2} & \multicolumn{1}{l|}{10.1} & \multicolumn{2}{c}{}\\ 
     \cline{1-3}
  \end{tabular}
\end{table}

%%%%%%%%%%%%%%%%%%%%%%%%%%%%%%%
%% TABLE 4 %%%%%%%%%%%%%%%%%%%
%%%%%%%%%%%%%%%%%%%%%%%%%%%%%%%
\begin{table}[t]
\caption{Test case II:  Notation as Table~\ref{table:exp0}.
%MAE (deg.) and Accuracy (\%) of VAE-SSL and alternative approaches. SRP-PHAT results in bottom row.
}\label{table:exp3}
  \centering
  \begin{tabular}{c | l | l | l | l }
    \cline{2-5}
     \multicolumn{1}{c|}{} & \multicolumn{2}{c|}{VAE-SSL} & \multicolumn{2}{c}{CNN}\\ \hline
     $J$ (\# labels) & MAE & Acc. & MAE & Acc.\\ \hline\hline
    37  & 28.5  & 9.63  & 50.0  & 8.90 \\
    74	& 13.0	& 31.7	& 43.5	& 15.3 \\
    148	& 16.0	& 36.6	& 33.2	& 21.0 \\
    481	& 6.19	& 65.9	& 11.5	& 53.8 \\
    999	& 4.28	& 74.4	& 0.85	& 92.1 \\
     \hline\hline
     \multicolumn{1}{c|}{SRP-PHAT} & \multicolumn{1}{l|}{22.0} & \multicolumn{1}{l|}{10.3} & \multicolumn{2}{c}{}\\ 
     \cline{1-3}
  \end{tabular}
\end{table} 

Thus in this case, we have 3 networks: 1) the label inference (classifier) network $\boldsymbol{\pi}_\phi(\mathbf{x})$ corresponding to $q_\phi(\mathbf{y|x})$; 2) the inference network $\boldsymbol{\mu}_\phi(\mathbf{x,y})$ and $\boldsymbol{\sigma}_\phi^2(\mathbf{x,y})$ corresponding to $q_\phi(\mathbf{z}|\mathbf{y,x})$; and 3) the decoder (generative) network $\boldsymbol{\mu}_\theta(\mathbf{y,z})$ corresponding to $p_\theta(\mathbf{x}|\mathbf{y},\mathbf{z})$. Graphical models representing the inference and generative networks are shown in Fig.~\ref{fig:ssvvae_dgm}. For optimization of the VAE-SSL system, the objective \eqref{eq:label_k6} is evaluated using probabilistic programming \cite{bingham2018pyro}.

\section{Experiments}

We assess the DOA estimation performance of the VAE-SSL approach in moderately reverberant environments against two alternative techniques: SRP-PHAT\cite{dibiase2001robust}, and a supervised CNN baseline. We simulate 4 room configurations to test the generalization of the learning-based methods in label-limited scenarios, as we expect this to closely approximate real applications. Thus, we test the learning-based methods (VAE-SSL and CNN) using few supervised labels, with $J$ the number of labels. The results are summarized in Tables~\ref{table:exp0}--\ref{table:exp3}, giving the performance of each method in terms of DOA error (mean absolute error, MAE) and frame-level accuracy. 

% We analyze the generated RTF from the VAE to help qualify the physics learned by the generative model, shown in Fig.~\ref{fig:reverb_exp}. 

All algorithms were coded in Python and the reverberant data was generated using Matlab \cite{habets2016}. The VAE-SSL system and CNN were implemented using Pytorch \cite{pytorch}, with the Pyro package \cite{bingham2018pyro} used for stochastic variational inference and optimization of the VAE-SSL. The NNs were optimized using Adam.

\begin{figure}[t]
\hspace{-1ex}
\centering
\includegraphics[width = 3.2in]{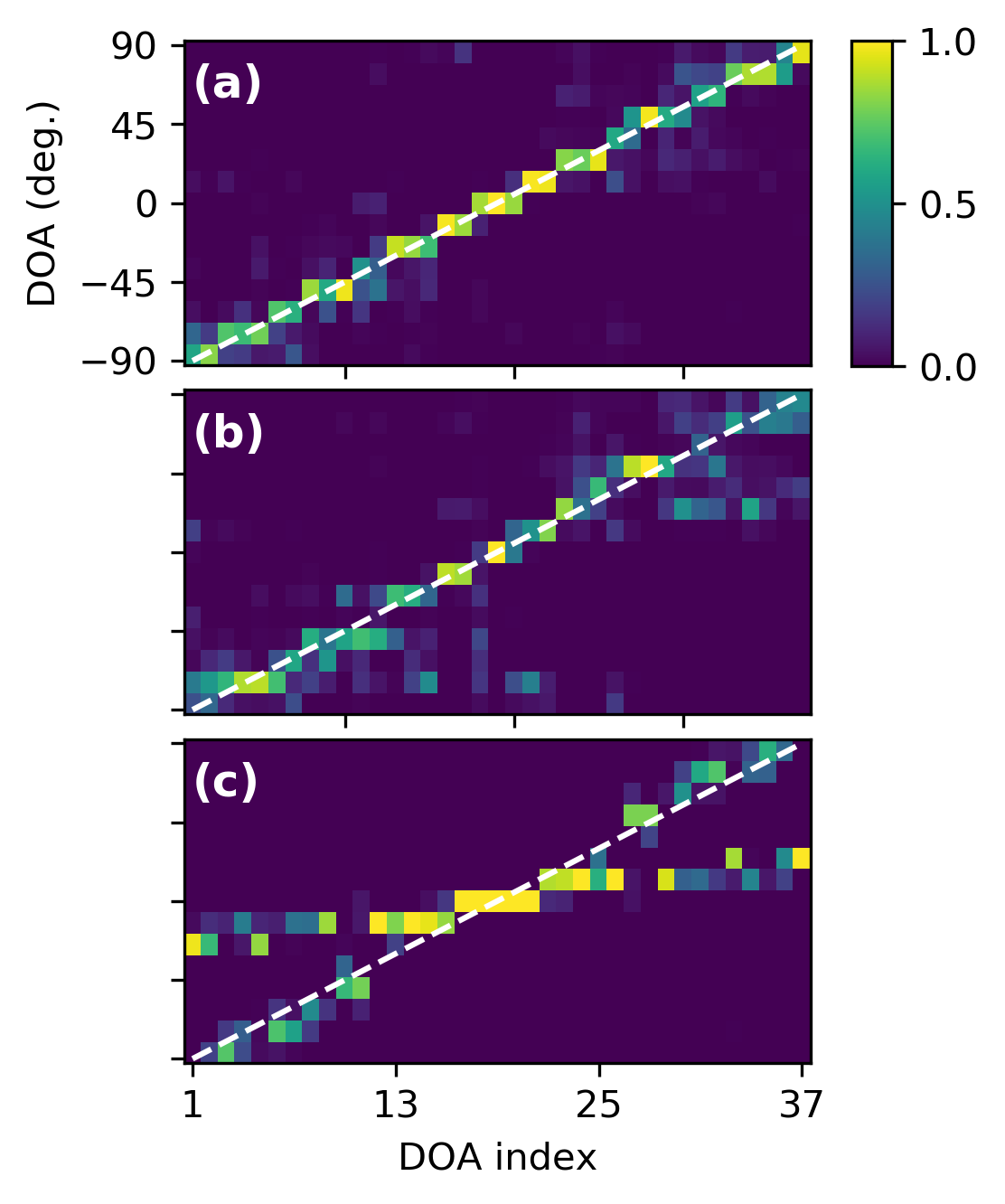}
\caption{Normalized histograms of DOA estimates by VAE-SSL and SRP-PHAT for each DOA index, for {\it Test case II}. VAE-SSL (a) $J=481$ and (b) $J=148$ labels. (c) SRP-PHAT results. True DOA is shown in each plot as white dashed line.}
\label{fig:loc_bins}
\end{figure}

\subsection{Reverberant room data}
\label{subsec:vae_noreverb}

The reverberant acoustic data were generated using the Room Impulse Response (RIR) generator \cite{habets2016}. The synthetic IRs are convolved with a Gaussian time-domain signal $s~\sim\mathcal{N}(0,1)$ (see \eqref{eq:mics}) and FFTs and RTFs were obtained for two sensors with only one active source location for each time bin. Noise signals were used for training/testing to avoid the difficulty of voice activity detection. We simulated 4 room configurations to test the generalization of the learning-based methods in label-limited scenarios. This included off-design conditions with changes to reverberation time and perturbations in the microphone positions.

The nominal room configuration, deemed the `design' case (see Table~\ref{table:exp0} for results) is a square room with x-y-z dimensions $6\times6\times2.4~\text{m}$ with a reverberation time $RT_{60}=500$~ms and $c=343$~m/s. Two omnidirectional microphones were located in the center of the room with spacing 8~cm. The source range was $1.5$~m, with sources at $5^\circ$ resolution for $[-90^\circ,90^\circ]$  azimuth relative to the array broadside ($T=37$ candidate DOAs). We used 1~s signals for each sensor location to obtain FFT and RTF frames with a 16~kHz sampling rate. A sensor noise level of 20~dB was assumed and 10 signal realizations were generated for each candidate DOA.

Three off-design rooms were simulated. The first, deemed the {\it validation} case (see Table~\ref{table:exp1}) had the same room configuration as the design case, but the reverberation time $RT_{60}=700$~ms. The second, deemed {\it test case I} (see Table~\ref{table:exp2}), had the room configuration and reverberation time as the design case, but the two microphone locations were displaced by 5~mm and 3~mm in opposite directions along the y-axis. The third, deemed {\it test case II} (see Table~\ref{table:exp3}), had the same room configuration as test case I, but $RT_{60}=600$~ms.

The signal at the microphones are given in \eqref{eq:mics}. We obtain the RTFs from the data by \eqref{eq:mics_rtf_fft2}. The RTFs are estimated using single FFT frames with Hamming windowing with 50\% overlap and $N_{FFT} = 256$. The VAE and supervised CNN inputs $\mathbf{x}_n$ use $P=32$ RTF vectors, giving an input size $K\times P= 32\times 128$. For fair comparison, SRP-PHAT used the $P=32$ FFT snapshots used to estimate the RTFs. Thus, the length of the analysis frame for all the methods was $0.26$~s. This yielded $\sim44,000$ (RTF) frames per room configuration.

\subsection{Learning-based model parameters}
The VAE-SSL model (classifier, inference, and generative networks) were implemented using CNNs. The classifier network $\boldsymbol{\pi}_\phi(\mathbf{x})$ and inference network ($\boldsymbol{\mu}_\phi(\mathbf{x,y})$ and $\boldsymbol{\sigma}_\phi^2(\mathbf{x,y})$) used the same architecture: two convolutional layers with 3-by-3 kernels with 2-by-2 max pooling, followed by two fully-connected layers with $200$ hidden units and output units. Each of the convolutional layers had 8 channels. The hidden units and convolutional layers used ReLU activation. The classifier network had $T$ output units with softmax activation. The inference network had $2\times M$ output units, corresponding to the latent code with dimension $M$, with linear activation. The convolution operations were applied to the RTF frames $\mathbf{x}$, and the labels $\mathbf{y}$ were input to the graph at the 200-unit hidden layer. The latent code dimension for all experiments was $M=2$, assuming that the hidden representation must account for source range (near-field) and temporal variation of the RTFs. 

The generative network ($\boldsymbol{\mu}_\theta(\mathbf{y,z})$) mirrors the inference network architecture, and consists of two fully connected layers of 200 and 2048 hidden units, followed by two transpose convolutional layers, each with a 2-by-2 max-unpool operations and transpose convolutions with 3-by-3 kernels. The hidden units and transpose convolution layer used ReLU activation. The output of the final layer had linear activation. Like the inference network, the labels $\mathbf{y}$ were input to the graph at the 200-unit hidden layer.

For comparison, we used the baseline `supervised-only' CNN from the VAE-SSL model, without unsupervised learning and integration with the generative model.

For all cases, the learning rate was 5e-5 and the batch size was 256. The default momentum and decay values (0.9,0.999) were used from the Pytorch implementation of Adam. The VAE-SSL has the tuning parameter, the auxiliary multiplier $\alpha$, which weights the supervised objective (see \eqref{eq:label_k6}). It was found in the experiments that the performance of VAE-SSL was not very sensitive to $\alpha$, though the best value per validation accuracy was found by grid search. 

\subsection{Non-learning: SRP-PHAT configuration}
The FFT features (used to calculate RTF) were used for the SRP-PHAT approach with the same number of frames as the VAE-SSL and CNN methods. SRP-PHAT used 37 candidate DOAs over $[-90^\circ,90^\circ]$. SRP-PHAT was implemented using the Pyroomacoustics toolbox \cite{scheibler2018pyroomacoustics}.

\subsection{Training and performance}

For VAE-SSL and supervised CNN training, $J$ labelled examples were drawn from the {\it design} case data to train the networks (i.e. calculate the losses and gradients) and $J$ labelled examples were drawn from the {\it validation} case data for validation. The model was chosen based on validation accuracy. For VAE-SSL, the additional unlabeled examples were used to train the networks. 

The full set of RTF features and labels for the experiments are $\{\mathbf{x}_n,\mathbf{y}_n\}\in\mathcal{D}$. From this set supervised samples with $\{\mathbf{x}_j,\mathbf{y}_j\}\in\mathcal{S}$ with $j$ the supervised sample index. The unsupervised examples (unlabeled) are $\{\mathbf{x}_l\}\notin\mathcal{S}$. All the RTF frames $\mathbf{X}=[\mathbf{x}_1\dots\mathbf{x}_N]$ were normalized to the range [0,1]. 

We consider a range of $J$ for assessing the performance of the DOA algorithms. The performance on the design, validation, test I, and test II cases (see Table~\ref{table:exp0}--\ref{table:exp3}) was assessed on the sample labels not available in training. We use $J=37,74,148,481,999$, which are multiples of the number of candidate DOAs ($T=37$) to ensure an equal number of labelled samples for each DOA. The case where $J=37$, corresponds to one labelled input frame for each DOA. For $J=999$, there are $27$ labelled input frames available per DOA.  

In VAE-SSL, for each value of $J$, the remaining samples are used for unsupervised learning, assuming no labels are available for those samples. During each training epoch, the supervised batches are used at a frequency proportional to their proportion of the overall data (supervised and unsupervised). Only the supervised samples were used to train the supervised CNN. Since for the unlabeled examples, the DOAs are technically unavailable, there are unsupervised frames $\mathbf{x}_l$ which contain a DOA change. For the supervised examples only input frames with a single DOA are used. 

The VAE-SSL auxiliary multiplier $\alpha$ values for each value of $J$ were chosen by grid search in the interval [10,100] in increments of 10. For $J=37,74,148,481,999$, the corresponding $\alpha=10, 50, 70, 20, 80$.

% \begin{figure}[t]
% \centering
% \includegraphics[width = 2.5in]{figures/cond_gen.jpg}
% \caption{{\color{red} \bf NEED TO UPDATE: (Top) RTF phase frames from the design case room configuration and (bottom) conditionally generated RTF phase using VAE-SSL model trained with $J=74$ labels. RTF samples are conditionally generated by $\widetilde{\mathbf{x}}\sim p_\theta(\mathbf{x}|\mathbf{y,z})$.}}
% \label{fig:reverb_exp}
% \end{figure}

The performance of the VAE-SSL method generally exceeds the supervised CNN and SRP-PHAT by a large margin for these label-limited scenarios, both in terms of MAE and accuracy. It is robust to small perturbations in sensor location and reverberation levels different from training. The supervised CNN outperforms the VAE-SSL approach for the maximum number of labels ($J=999$). However, as the number of labels decreases, the performance of VAE-SSL exceeds that of the supervised CNN and SRP-PHAT. VAE-SSL outperforms SRP-PHAT, both in terms of MAE and accuracy for as few as two labeled input samples for each DOA ($J=74$). The performance of VAE-SSL is compared with SRP-PHAT in Fig.~\ref{fig:loc_bins} for only few labelled examples. In general, SRP-PHAT was outperformed by the learning-based methods. 

% \subsection{Generating RTFs with VAE-SSL}

% The trained VAE-SSL can conditionally generate RTF phase. We generate the reverberant RTF phase using generative network and plot it relative to the labelled input in Fig.~\ref{fig:reverb_exp}. RTF samples are conditionally generated by $\widetilde{\mathbf{x}}\sim p_\theta(\mathbf{x}|\mathbf{y,z})$, holding $\mathbf{z}$ constant, and sampling over the DOA labels $\mathbf{y}$. We use the phase-wrap of the RTF ($\pm\pi$) (as a function of sensor separation and DOA $\theta_n$, $f=c/(2r|\sin\theta_n|)$) to help qualify the physics learned by VAE-SSL. This is plotted along with the RTF frames from the design case room configuration. It is observed that the physics of the RTF phase are well-learned by the VAE-SSL model.

\section{Conclusions}
This study shows that deep generative modeling can   localize sources well in reverberant environments, relative to existing approaches, when only few labels are available. It was  demonstrated for moderately reverberant scenarios that VAE-SSL outperforms supervised CNN when only few labels are available, provided significant unlabeled data is available. Training the VAE-SSL on all reverberant frames (with and without labels) allows the VAE system to exploit the structure of the data. 
%The strength of the VAE-SSL approach lies in learning from both labeled and unlabeled examples is evident. 
The experiments show, only two labeled samples per DOA  permit the VAE-SSL to obtain better performance than SRP-PHAT. 

% Further, the representations learned by such generative approaches can be used to generate new samples.

\small
%\bibliographystyle{IEEEbib}
%\bibliography{biblio_gen_doa0.bib}

\end{document}